\documentstyle[12pt,aps,eqsecnum]{revtex}
\begin{document}
\draft
\title{Optimal self-avoiding paths in dilute random medium}
\date{\today}
\author{ F.Seno$\null^1$, A.L. Stella$\null^2$
 and C. Vanderzande$\null^3$ }

\address{(1) INFM-Dipartimento di Fisica, Universit\`a di Padova, Italy }

\address{(2) INFM-Dipartimento di Fisica e Sezione INFN, Universit\`a
di Padova, Italy }

\address{(3)  Department Wiskunde Natuurkunde Informatica, Limburgs
Universitair
Centrum, 3590 Diepenbeek, Belgium}
\maketitle
\begin{abstract}

By a new type of finite size scaling analysis on the square lattice, and by
renormalization group calculations on hierarchical lattices we investigate the
effects of dilution on optimal undirected
self-avoiding paths in a random environment. The behaviour of the
optimal paths
remains the same as for directed paths
in undiluted medium, as long as forbidden bonds
are not exceeding the percolation threshold.
Thus, overhanging configurations do not alter
the standard self-affine directed polymer scaling  regime, even above
the directed threshold, when they become unavoidable.

When dilution reaches the undirected threshold, the
optimal path becomes fractal, with fractal dimension
equal to $D_{\rm min}$, the dimension of the minimal
length path on percolation cluster backbone.

In this regime the optimal path energy fluctuation, $\overline{\Delta E}$,
can be ascribed entirely to minimal length
fluctuations, and satisfies $\overline{\Delta E} \propto L^{\omega}$,
with $\omega=1.02 \pm 0.06$ in $2d$, $L$ being the Euclidean
distance. Hierarchical lattice calculations confirm that $\omega$ is
also the exponent of the leading scaling correction to
$\overline E \propto L^{D_{\rm min}}$.
 Upon approaching threshold, the
probability, ${\cal R}$, that the optimal
path does not stick entirely on the
minimal length one, obeys ${\cal R} \sim \left( \Delta p\right) ^{\rho}$,
with $\rho \sim 1.0 \pm 0.05$ on hierarchical lattices. Such behaviour
could be characteristic of the crossover to fractal regime.
Transfer matrix results on square lattice show that a similar full
sticking does not
occur for directed paths at the directed percolation threshold.

\end{abstract}

\newpage

\section{Introduction}

The problem of directed paths in a random potential has attracted
a lot of interest because of its connection with issues like domain walls
in random ferromagnets\cite{HHb85}, interface growth\cite{KPZ}, fracture
and failure phenomena\cite{HHR} and magnetic flux lines in high-temperature
ceramic superconductors\cite{Nel90}.

At zero temperature the interface separating oppositely magnetized phases in a
$2d$ random Ising ferromagnet is given by a minimal energy
path on the dual lattice. Upon averaging over randomness the path has
self-affine geometry with roughness exponent $\zeta=2/3$\cite{HHb85}.
 The energy
fluctuates about
average with a longitudinal size dependence described by the exponent
$\omega=1/3$\cite{HHb85,KPZ}. Up to now, models exhibiting this kind of
behaviour have always been
studied with the simplifying restriction of allowing only
directed path configurations ( directed polymer in random medium: DPRM). This
means that the path, besides being self-avoiding, can not
develop overhangs. Such a restriction is not expected to be
essential for obtaining a path with self-affine geometry at large
length scales. On the other hand, removal of the directedness constraint on
the path must be crucial, e.g., for those situations in which
disorder becomes of such a nature to induce crossover from self-affine
to fractal geometry.

In this paper we address the problem of determining the scaling
properties of optimal (undirected) self-avoiding paths, under the effect
of a percolative,
geometrical disorder which forbids a given fraction of lattice edges
to the path. Such disorder, associated to the existence of percolative voids,
is considered in combination with the usual energy randomness of the
remaining accessible bonds. Models of this type, but directed, were considered
recently by Balents and Kardar \cite{BK92} and are expected
to have applications to fractures of porous materials\cite{RF91}
and interface growth of poisoned Eden models\cite{CW91,KS86}.

In the context of domain walls in random ferromagnets we can imagine a $2d$
Ising problem in which a fraction, $p$, of nearest neighbour spins
has an extremely strong, essentially infinite, ferromagnetic exchange,
while the others have a distribution of random couplings with finite
limits and variance.

At $T=0$ the domain wall separating oppositely magnetized phases will of course
not pass through the duals of infinite exchange bonds, which as a result are
excluded from the wall.

 One should notice
that, for a configuration of exchange strengths in which walls
fully excluding infinitely strong couplings are not allowed,
the minimal energy of the domain wall would be infinite as well. Our choice
is to exclude such case from the statistics. So, the optimization
problem looses meaning for us in case $p$ exceeds the percolation
threshold. An operative way to realize the exclusion could consist in using
appropriate boundary conditions, able to push the interface to stick on
one of the edges of the sample as soon as the otherwise optimal path
should cross infinitely strong bonds.

 Another case in which the optimal interface has to develop within
a restricted space is when the system undergoing a phase transition
is hosted in the pores of inert structures like aerogels. A possibility
of describing these situations is by means of spin models in a suitably diluted
and random environment \cite{UHJ95}.

For the above interfaces, the optimal configuration is thus that in which
energy is minimized by passing only through duals
of the weakest among accessible random bonds, which amount to a fraction
$1-p$ of the total.

In ref \cite{BK92} it was discovered
that a directed path in such an environment remains of the standard type,
with $\zeta=2/3$ and $\omega=1/3$, as long as $(1-p) > p_{\rm cd}$, where
$p_{\rm cd}$ is the directed percolation threshold for allowed bonds. Right
at $(1-p)=p_{\rm cd}$ the optimal path has to pass through the incipient
infinite cluster ($IIC$) of directed percolation and its properties change
drastically. Indeed one finds \cite{BK92} $\zeta = 0.61 \pm 0.04$
and $\omega\sim 0.50 \pm 0.01$, which
suggest $\zeta=\nu_{\perp}/\nu_{\parallel}$, $\nu_{\perp}$ and
$\nu_{\parallel}$
being the correlation length exponents of directed percolation in
$2d$\cite{Kinzel83}.

A main issue related to such a transition can be formulated as follows.

Suppose we release the constraint of directedness for the paths.
We know that, in this case, the directed percolation threshold marks only
the point below which one can not find anymore $\underline{directed}$ paths
 within clusters of allowed bonds. A lower threshold,
$p_c < p_{\rm cd}$, marks now the complete
disappearance of connecting paths with overhangs. It is interesting
to investigate the nature of the optimal path for $p_c < (1-p) < p_{\rm cd}$.
Indeed, this is a situation in which $\underline{only}$ overhanging
 configurations
are allowed by the disorder, and one can not exclude new scalings to
arise. Indeed the possible effect of overhangs exclusion on asymptotic
behaviour is a very debated and controversial issue, especially in the context
of interface growth models, which are closely related to DPRM\cite{KS94}.

A further interesting problem is that of determining the precise behaviour
of the optimal path when it has to pass through the backbone of the
(undirected) $IIC$, which is a fractal structure. It is of particular
interest to determine the new scalings characterizing geometry
and energetics of the path also in this condition. Besides establishing how the
laws of the self-affine regime generalize to the fractal one, such a study
should
also elucidate the possible connection between path scaling properties
and backbone geometry.

As we will  discuss below, clear cut answers to the above and other, related
questions are often very difficult to give if one works only with paths on $2d$
regular lattice. The reason is that, as soon as the walks are
not directed, the standard statistical treatment, based
on the transfer matrix, becomes extremely time and memory
consuming and does not allow easy extrapolation of finite size results.

This is the main reason why, in the recent past, most
of the issues we are facing here were essentially not addressed.

In this paper we produce a main effort towards implementation of
a finite size scaling (FSS) analysis based on transfer matrix
of self-avoiding paths in geometrically and energetically random
$2d$ environments.

At the same time, in view of the above mentioned difficulties,
we try to gain additional insight by an extensive investigation
of realizations of our models on suitable hierarchical lattices.
On such lattices one has the great advantage that renormalization
group (RG) recursion equations can be written exactly.
By numerical methods, such equations can be iterated long enough
to allow satisfactory control of asymptotic scaling
properties.

The results we present for hierarchical lattices should be considered as
important and substantial complements, and sometimes substitutes, of findings
in the Euclidean lattice.

The present paper is organized as follows.
In the next section,
we introduce  our model on Euclidean lattice and its hierarchical
realizations. Renormalization group results for hierarchical lattice are
also presented and discussed there. The third section is devoted
to the transfer matrix approach on square
lattice. Results concern both undirected and
directed self-avoiding walks (SAW).
A summary of the results, together with concluding remarks are
given in the last, fourth section.

\section{ Optimal self-avoiding paths on hierarchical lattices.}

As anticipated in the introduction, we want to consider a path
optimization problem which can be formulated as follows in the
case of a $2d$, square lattice. With an independently and
identically distributed probability,
$p$, each lattice edge can be assigned the role of inaccessible
(e.g., infinite energy) bond in a given random configuration. To the remaining,
accessible bonds, $b$, finite positive energies $E_b$ are assigned
according to an independent distribution ${\cal P}(E_b)$. For
each random configuration obtained in this way we then try to
determine the minimal energy self-avoiding paths (if any) connecting
(through accessible bonds) two given points at distance $L$. The energy
of a self-avoiding path $W$ is of course given by

\begin{equation}
E_{W}= \sum_{b \in W} E_b
\label{eq:2_1}
\end{equation}

Of particular interest is to determine the optimal path energy fluctuations
with respect to the average over all random configurations. Such fluctuations
can be described by

\begin{equation}
\overline{\Delta E} =
\left( \overline{
(
E_{\rm min}-\overline{E_{\rm min}}
)^2   } \right)^{1/2}
\label{eq:2_2}
\end{equation}

where, in a given configuration,

\begin{equation}
E_{\rm min} = {\rm \min}_{W} \left( \sum_{b \in W} E_b \right)
\label{eq:2_3}
\end{equation}
and the bar indicates average over all configurations.

In the case of directed paths in random, undiluted medium,
 it is found that
$\overline{\Delta E} \sim L^{\omega}$ for $ L \to \infty$ \cite{HHb85}. This
scaling is related to the geometrical behaviour of the path
in a simple way. Indeed, for directed paths, this geometry
is always self-affine, with transverse fluctuations with respect
to the straight path of length $L$, growing as $L_{\perp} \sim L^{\zeta}$,
$\zeta$ being a roughness exponent. By linking energy to length
fluctuations\cite{HHb85}
one can also derive

\begin{equation}
\omega = 2 \zeta -1
\label{eq:2_4}
\end{equation}

The $\omega$ exponent determines also the leading scaling correction
to the average optimal path energy, according to

\begin{equation}
\overline{E_{\rm}} \sim L (1+cL^{\omega -1}) \ \ \ {\rm for}\ \  L \to \infty
\label{eq:2_5}
\end{equation}

In ref. \cite{BK92} the same problem with dilution
 defined above, but restricted  to
directed paths, was considered at finite temperature. In
that case each walk $W$ is weighed by a Boltzmann factor
$\exp{\left( -E_{W}/T \right)}$, and all admissible walks enter in the
statistical averages
for any given configuration. Rather convincing
evidence was obtained in ref \cite{BK92} that, as
long as $ (1-p) > p_{\rm cd}$, the directed paths behave
with exponents $\omega=2/3$ and
$\zeta=2/3$ satisfying eq. (\ref{eq:2_4}). When $(1-p)=p_{\rm cd}$, the
behaviour is still self-affine, but with different exponents
$\omega = 0.50 \pm 0.01$ and $\zeta = 0.61 \pm 0.04 \sim
\nu_{\perp}/\nu_{\parallel}$,
$\nu_{\perp}$ and ${\nu_{\parallel}}$ being the correlation
length exponents of directed percolation\cite{Kinzel83}.
In this case eq. (\ref{eq:2_4}) is
not satisfied anymore, and the result for $\zeta$ can be understood in the
sense
that, for $(1-p)=p_{\rm cd}$, the path is limited by the
backbone of the directed percolation $IIC$. This
limitation of an otherwise standard walk ($\omega=1/3, \zeta=2/3$) has been
indicated as possibly responsible of the violation of eq. (\ref{eq:2_4})
\cite{BK92}. Our goal in this work is to understand how all this possibly
modifies when considering undirected optimal paths, and to describe the new
physics setting in when dilution is pushed further than directed percolation
threshold.

In the next section we illustrate several results obtained by applying transfer
matrix methods to our undirected model and to a variant of that treated in
ref. \cite{BK92}.

However, the difficulty of the subject suggests to approach the main
physical issues also in the context of hierarchical models. Indeed,
for these models the RG approach can give quite accurate results
at a very reasonable computational cost. If not a good approximation,
such results should at least provide a rather complete, qualitative picture
of the behaviour of models on Euclidean lattice. This is of outmost importance
in our case.

Let us consider a hierarchical lattice obtained by following the iteration
scheme illustrated in Fig. 1a. The resulting lattice is a diamond with
a diagonal bond (DDHL). So far, diamonds without
diagonal (DHL) (Fig. 1b) have been always considered
for the description of directed paths in random media\cite{Stella94}.
 However, on a DHL
all paths joining the extrema have the same length ($2^{n}$) at the
$n$--th stage of construction, and there is no way of introducing a distinction
between ``directed" and ``undirected" paths on them.

Here a self-avoiding path joining the DDHL ends will be considered
directed if, by following its development through the various stages
of the construction, it never passes through the diagonal of
any diamond. In other words, directed paths on DDHL are those which could be
hosted by a DHL. As a matter of fact the DDHL was already
used recently to represent
the properties of SAW in random environment\cite{DM91}. While this
reference focuses on the properties in the regime corresponding
to the standard criticality of SAW on regular lattice, here we
are concentrating on what can be called the
stretched regime, and on the effects of directedness removal on it.

 On a DDHL the directed percolation threshold is at
$(1-p)=p_{\rm cd}= {(\sqrt{5}-1)/2}$. The undirected threshold is at $(1-p)=
p_c=1/2$. For the renormalization group treatment we consider the joint energy
and length probability distribution for optimal paths crossing the
DDHL at stage $n$. If $N(W)=\sum_{b \in W} 1 $ is the number of steps
of $W$, we indicate this probability by
${ P}_{n}(E,N)$. With our initial condition clearly:

\begin{equation}
{ P}_{0}(E,N) = \delta_{N,1} [(1-p){ P}(E) + p \delta(E - \infty)]
\label{eq:2_6}
\end{equation}
where ${\cal  P}$ is the energy probability distribution of accessible bonds.
With our assumption that no bond with infinite energy can be crossed by
the optimal path, averages based on eq. (\ref{eq:2_6}) must of course
ignore the second term in brackets, and include division by $(1-p)$ in order
to restore normalization.

${ P}_{n+1}$ can be constructed from ${ P}_{n}$ as follows:

\begin{equation}
{ P}_{n+1}(E,N)=
\sum_{\{N_{\alpha}\}} \int \Pi_{\alpha=1}^{5}
\left[
dE_{\alpha}{ P}_{n}(E_{\alpha},N_{\alpha}) \delta
(E-{\rm \min}_W \sum_{\beta \in W} E_{\beta})
\delta_{N,\sum_{\gamma \in W_{\rm min}} N_{\gamma}}
\right]
\label{eq:2_7}
\end{equation}
where $\alpha$ numbers the five bonds of a diamond cell with diagonal (Fig.1a),
$W$ are the $4$ distinct self-avoiding paths
crossing  the cell, and $W_{\rm min}$ indicates among them
that for which $\sum_{\beta \in W} E_{\beta}$ is minimal.

Of course, also ${ P}_n$ will contain a part proportional
to $\delta(E-\infty)$. By construction, the coefficient of this term,
which has to be excluded in the discussion of the allowed optimal paths,
is proportional to $p^{(n)}$, the $n$--iterated of the RG transformation

\begin{equation}
p^{\prime}=(2p^3-5p^2+2p+2) p^2
\label{eq:2_8}
\end{equation}
for bond percolation on the DDHL.

Thus, for $p< p_{c}$, the recursion (\ref{eq:2_7}) drives
${ P}_n$ towards $p=0$, i.e. to a situation where excluded, $E=\infty$,
bonds disappear. For $ p > p_{c}$, only excluded bonds survive for
$ n \to \infty$, and no optimal path exists, in the sense explained
above. For $p=p_c=1/2$, the concentration of forbidden bonds remains
fixed under iteration and only their energy distribution evolves.

Iteration of eq. (\ref{eq:2_7}) can not be performed exactly. A simple and
accurate numerical strategy consists in iterating a large sample
of paths distributed in energy and length according to ${ P}_n$.
In the interest of easier computational tractability, we choose

\begin{equation}
{\cal P}(E)=q\delta(E-1)+(1-q)\delta(E-2)
\label{eq:2_9}
\end{equation}
Our normal choice will be $q=1/2$, in case we want to treat a finite width
distribution. $q=0$ will apply to the case of zero width. As far as
$q=1/2$ is concerned, we expect, and explicitly verified, that
 different choices of $q$ or of the energy values are not relevant.
We expect the $q=1/2$ asymptotic results to belong to the same universality
class as those obtained with more standard distributions on continuous
interval. Our ${\cal P}$ offers the computational advantage that optimal path
energies remain positive integers.

The initial sample of paths is a large collection of
$(E_{\rm i}^{(0)},N_{\rm i}^{(0)}), i=1,2, \ldots i_{\rm max}$
$(i_{\rm max} \leq 10^6)$, randomly generated according to
(\ref{eq:2_6}). Groups of $5$ pairs are extracted from the sample,
and each couple is associated to
one of the $5$ bonds of the DDHL cell. For each group the optimal
path across the cell is identified, and its total energy, and length are stored
in the new ensemble as $(E_{\rm i}^{(1)},N_{\rm i}^{(1)})$, and are distributed
according  to ${ P}_{1}$. This can be iterated several times
before serious statistical problems due to limited sampling arise.

The knowledge of ${ P}_n$ allows to determine
quantities like $\overline{E_{\rm min}}$, $\overline{\Delta E}$ or the average
path length, $\overline{N}$, as a function of the initial conditions
and of $L=2^n$, the (longitudinal) length of the lattice. We could
normally handle $n \leq 20$ in our calculations. Of course the procedure
just described can be applied also to the same model on DHL, or
on
more complicated lattices. The percolation threshold for DHL,
$(\sqrt{5}-1)/2=0.6180\ldots$, coincides with $p_{\rm cd}$ for DDHL.

For our model on DDHL we first considered the case $(1-p)>p_{\rm cd}$, with
energy disorder specified by $q=1/2$. The asymptotic regime here is not
modified, with respect to the DHL case, by the presence
of undirected configurations. Indeed we find $\omega=\log(1.230 \ldots)/\log(2)
\sim 0.30$, which coincides, within the numerical accuracy,
with the value first obtained in ref \cite{DG89}. On DHL this
$\omega$ nicely
corresponds to the DPRM value $\omega=1/3$ on Euclidean lattice.

Assuming validity of (\ref{eq:2_4}) one gets also a
 roughness exponent $\zeta\sim0.65$, obviously
close to the DPRM value 2/3.
This result, for $(1-p)>p_{\rm cd}$ is certainly consistent with what is
expected to happen on Euclidean lattices, but was never explicitly verified
so far, to our knowledge. At $(1-p)=p_{\rm cd}$ the path on DDHL
behaves differently from that on DHL. Indeed, for DHL, $(1-p)=p_{\rm cd}$
is the percolation threshold, below which the optimization problem
looses meaning. One can verify that, at
$(1-p)=p_{\rm cd}$, $\omega \sim 0.49 \pm 0.02$ on DHL,
whereas $\omega \sim 0.30$ remains for DDHL.

$\omega \sim 0.50$ was already found for $T>0$ on DHL in ref. \cite{BK92},
where it was shown to be in agreement with the transfer matrix result
for directed paths on square lattice, $\omega =0.50 \pm 0.01$.

Our finding for DDHL shows that the directed threshold is not felt at all
by the undirected optimal path. Moreover, we find that, in the
whole region $p_{c} < (1-p) \leq p_{\rm cd}$, $\omega$ keeps the value $0.30$.
 This means that, even if optimal path
configurations must now contain overhangs, these are not able
to affect the large scale behaviour. A further confirmation
of the fact that the optimal path on DDHL has to be considered
self-affine  for $p_c<(1-p) \leq 1$, comes from a computation
of the fractal dimension, $D$, of the path, based on $\overline{N} \sim
(2^n)^D$.
With very high accuracy we find $D=1$ on the whole range. Notice that on
DHL, which in our context represents the implementation of the directedness
constraint, $D=1$ holds by construction.

 We conjecture therefore that a self affine DPRM behaviour with $\omega=1/3$
and $\zeta=2/3$  should hold on Euclidean
lattice as long as $(1-p)>p_c$.


Of particular interest is the case $(1-p)=p_{c}=1/2$ on DDHL. For
this threshold case we find $\omega=1.04 \pm 0.05$. Assuming validity of
 eq. (\ref{eq:2_4}), $\omega=1$ would correspond to $\zeta =1$, the
limiting value where self-affinity changes into fractality.

The fractal character of the path is demonstrated by
$D = 1.085 \pm 0.005$. In fact the optimal path at threshold is
constrained to develop within the backbone of the infinite incipient
percolation cluster, and we can try to identify
$D$ among the fractal dimensions of the backbone on DDHL.
A natural candidate is the dimension $D_{\rm min}$ of the minimal
length path across the backbone. One way to test this is that of
comparing $D$ values for disordered bond energies ($q=1/2$),
and for a case with no disorder ($q=0$, for example). In the latter
case, the optimal path has to coincide with the minimal length path,
thus $D = D_{\rm min}$ must hold. Remarkably, at $1-p=p_c$ we find no
appreciable
differences in the $D$ and $\omega$ values for $ q=1/2$, and
$q=0$. The fact that also $\omega$ remains unchanged by putting $q=0$,
means that energy fluctuations in this regime are  dominated by the
geometrical backbone disorder: thus
$\omega$ has the geometrical meaning of describing the fluctuations
of minimal chemical distance within the backbone.
The presence or absence of energy disorder in the backbone bonds does not
affect
the value of $\omega$, which is just a direct manifestation of
the $IIC$ geometry.

A further confirmation that $D=D_{\rm min}$ comes from a direct exact RG
evaluation of $D_{\rm min}$ on DDHL, which gives
$D_{\rm min}=\log{(17/8)}/\log{2}=1.089$.
This determination can be done straightforwardly by appropriately
weighing, at $p=p_c$, the lengths of all possible path crossing a percolating
cell and performing the average.

At first sight
$\omega=1$ seems to qualify as a natural exponent for optimal paths
having a fractal geometry in view of the relation (\ref{eq:2_4}). Notice,
however,
that the validity of eq. (\ref{eq:2_4}) is not a priori obvious: at the
directed
percolation threshold directed paths on Euclidean lattice do
not satisfy such relation, for example.

Also as an indication of the predictive value of this kind of model
calculations, it is worth checking the robustness of $\omega$
with respect to modifications of the hierarchical lattice. We used
other lattices, with more complicated cells, reproducing
features of the square lattice and allowing
distinction between directed and undirected thresholds.
Two examples are reported in Fig.  1c and Fig. 1d.

 With similar, somewhat more heavy RG calculations, we could confirm
the scenario described above for the DDHL. In  particular,
at the percolation thresholds we could determine $\omega=1.02 \pm 0.05$ and
$\omega = 1.01 \pm 0.05$, rather consistent with the DDHL result.
We should expect the results for $\omega$ in the case of the lattices in
fig. 1c and 1d  to represent better approximation of the Euclidean exponent.

To better elucidate the meaning of $\omega$ in the case of
fractal optimal  path, we also checked how eq. (\ref{eq:2_5}) changes
in this case. Our data for $\overline{E_{\rm min}}$ could be fitted
very well by the form:

\begin{equation}
\overline{E_{\rm min}}(L) \sim a L^{D_{\rm min}} + b L^{\omega} \ \ \ L \gg 1
\label{eq:2_10}
\end{equation}
with $D_{\rm min}=1.08$ and, $\omega=1.01 \pm 0.03$ and $L=2^n$.
Eq. (\ref{eq:2_10}), is consistent with
an asymptotic distribution

\begin{equation}
Q_n(E)=\sum_N { P}_n(E,N)= \frac{1}{L^{\omega}}f\left(
\frac{E-E_{\rm min}(L)}{L^{\omega}} \right)
\ \ \  (E < \infty)
\label{eq:2_11}
\end{equation}
of optimal paths energie widths. Thus $\omega$
should be the only scaling dimension for energy.

A stressed above, the fact that, at $(1-p)=p_c$, $D$ and $\omega$ remain
the same for $q=0$,
 indicates that $D=D_{\rm min}$, and that energy fluctuations must be fully
ascribed to
optimal path length fluctuations.
In other words, at large scales, the optimal path must
stick completely
on the minimal
length path.

It is interesting to contrast this behaviour with the
situation at $(1-p)>p_c$.
To this purpose one can introduce a quantity ${\cal R}_n$, which represents
the probability that, on a DDHL of length $2^{n}$, the optimal path deviates
from the minimal length one. This quantity can be easily computed
by checking, at each RG iteration, the fraction
of optimal paths for which the choice of minimizing $E$ is not leading to a
simultaneous minimization of $N$. ${\cal R}_n$ can be easily followed in
different regimes. We find that, as long as $(1-p)>p_c$, ${\cal R}_n$
approaches
a nonzero limiting value for $n \to \infty$.This indicates that in
the self-affine scaling regime deviations from the minimal path always occur.
On the other hand,the limiting value of ${\cal R}_n$
appears to approach $0$
for $(1-p) \searrow p_c$,
consistent with what we found above.  A fit of the behaviour of $ {\cal R}
\equiv
\lim_{n \to \infty} {\cal R}_n$ as a function of $(1-p)$ gives:

\begin{equation}
{\cal R} \simeq [1 -p -p_c]^{\rho} \ \ \ {\rm for} \ \ (1-p) \searrow p_c
\label{eq:2_12}
\end{equation}
with $\rho=1.0 \pm 0.05$. This is a new critical exponent associated to
the crossover from self-affine to fractal regimes.

Similar values of $\rho$ could be estimated for the lattices in Fig. 1c and 1d.
The quantity ${\cal R}$ with its behaviour (\ref{eq:2_12}) offers a further
characterization of the transition from
self-affine to fractal regime.
This transition can be interpretated as a full adsorption undergone by the
polymer on the minimal length path. It would be quite interesting to
obtain an estimate of $\rho$ for paths on Euclidean lattice. Unfortunately
such an estimate did not turn out to be feasible with the methods we describe
in the following section.

Another interesting issue is whether a quantity like $\cal R$ behaves in
similar way for directed walks at the directed percolation threshold.
Unfortunately, since on DHL all paths have the same length, a computation of
 ${\cal R}$ on DHL for directed paths does not make
sense . However, in
the next section
we compute this quantity for directed paths on square lattice ( direction
along one of the lattice axes and solve the issue).

\section{Finite size scaling approach to optimal
self-avoiding paths on 2d square lattice}

We now turn to a study of self-avoiding paths on an Euclidean
square lattice. We call $x$ the
abscissa and $y$ the ordinate. The paths have one end fixed in \
$(0,0)$. As in the previous section, each bond can be accessible or not,
and to the accessible bonds random energies are associated,
consistently with (\ref{eq:2_6}). In our numerical calculations we will
again use (\ref{eq:2_9}) with $q=1/2$.

For a given $x=t$ we consider all possible paths joining the
origin with the $x=t$ vertical axis (we ``cut" the paths at the
first intersection with this axis, i.e. the intersection at
shortest `` chemical" distance from the origin).

For each $y$ on the $x=t$ axis we consider the minimal energy
of the paths arriving at $y$, $E_{\rm min}(t,y)$. The minimal
energy fluctuations can be estimated by\cite{KZ87}:

\begin{equation}
\Delta E (t) = E_{\rm min}(t,0)-E_{\rm min}(t)
\label{eq:3_1}
\end{equation}
where

\begin{equation}
E_{\rm min}(t)= {\rm \min}_y\left\{E_{\rm min}(t,y)\right\}
\label{eq:3_15}
\end{equation}
and $y_{\rm min}(t)$ indicates the $y$ at which the minimum energy is reached.
  In the diluted case ($(1-p)<1$) the deviation (\ref{eq:3_1}) could
be meaningless in case no path arrives at $(t,0)$. To avoid
this one can choose, e.g., the point closest to the origin
instead of $(t,0)$ in each configuration.
Alternatively, as in the previous section,
one can take for $\Delta E (t)$ the root mean square variance
over all $E_{min} (t,y)$.
The general ideas on path optimization, and the results of the previous
section let us expect the following scalings:

\begin{equation}
\overline{\Delta E}  \sim t^{\omega} \ \ \ t \to \infty
\label{eq:3_2}
\end{equation}

\begin{equation}
\overline{\mid y_{\rm min}(t) \mid} \sim t ^{\zeta} \ \ \ t \to \infty
\label{eq:3_3}
\end{equation}
where again the bars indicate averages with respect to random
bond configurations.

So far, we assumed that the self-avoiding paths can occupy the whole
half-plane $0 \leq x, -\infty < y < + \infty$.

Of course it is hopeless to deal with all such paths. In order to determine
the exponents in (\ref{eq:3_2}) and (\ref{eq:3_3}) we propose
here a finite size scaling (FSS) strategy which exploits the possibility of
considering
the problem restricted to rectangular boxes, $ 0 \leq x \leq t, 1\leq y
\leq L$, with $L$ finite (and relatively small), and $t$ ranging up
to values much larger than $L$. Optimal self-avoiding path within such boxes
 should obey finite size version of (\ref{eq:3_2}) and (\ref{eq:3_3}).
  For example

\begin{equation}
\overline{\Delta E(t,L)}
 \sim t^{\omega}F\left(
{t^{\zeta} \over {L}}
\right )
\ \ \ t \to \infty
\ \ \ L \to \infty
\label{eq:3_4}
\end{equation}
$F$, consistently with eq.(3.2), should satisfy

\begin{equation}
\lim_{z \to 0} F(z) ={\rm const}
\label{eq:3_5}
\end{equation}
and, in view of the fact that, at fixed $L$ and for
$t^{\zeta} \gg L$, ${\overline{\Delta E}}$ should approach a $t$ independent
saturation value,

\begin{equation}
F(z) \sim z^{-{\omega}/{\zeta}}
\label{eq:3_6}
\end{equation}

This property follows from the fact that $\overline{\Delta E}$ is defined
as average over bond configurations of a quantity which is
a $\underline{difference}$ between $\underline{two}$ energies
in each configuration. If, on the contrary, we would define
$\overline{\Delta E}$ as the root mean square variance of $E_{\rm min}(t)$
over all configurations we would get $\overline{\Delta E}(t,L)
\sim t^{\frac{1}{2}}$. This last result is a consequence of the central limit
theorem. Indeed, for $t^{\zeta} \gg L$, the box becomes essentially
one-dimensional and $E_{\rm min} (t)$ can be seen as the sum of many
independent random terms, corresponding to different segments along the
$L$ axis.

Eq. (\ref{eq:3_6}) can be used to estimate $\omega$ and $\zeta$ by
extrapolating results for
relatively small $L$. If we call $N(t)$ the number of steps of the
minimal energy path within the rectangle $0\leq x\leq t,1\leq y\leq L$
one expects the following FSS scaling form for the averaged $N(t)$

\begin{equation}
\overline{N(t)} \sim t^{D} G \left({t^{\zeta}\over L}\right)
\label{eq:3_8}
\end{equation}
where
$G(z)$ approaches a constant for $z$ approaching zero,
so that $\overline{N(t)} \sim t^{D}$ for $ L \gg t^{\zeta}$.

On the other hand, for $t^{\zeta} \gg L$, we must
have $\overline{N(t)}\sim t$ \footnote{ This follows from the fact that
$ t \leq N(t) \leq Lt$ and thus must be linear in $t$ for
large $t$.}, so that $G(z) \sim z^{ \frac{1-D}{\zeta} }$.

Thus, for $ t^{\zeta} \gg L \gg 1$

\begin{equation}
\overline{N(t)} \sim t L^{ \frac{D-1}{\zeta} }
\label{eq:3_9}
\end{equation}

In case the path is fractal, we expect $ D > 1$ and $\zeta = 1$, while for
the self-affine case $D=1$ and $\zeta < 1$, in eq. (\ref{eq:3_9}).
As a consequence $\overline{N(t)} \sim t L^{D-1}$ in the fractal
regime and $\overline{N(t)} \sim t$ in the self-affine regime.
The behaviour of $\overline{N}$ for $t^{\zeta} \gg L$ thus allows a direct
estimate
of $D$.

For each bond configuration in a box the optimal
self-avoiding path energy and its length
have been computed
exactly by a transfer matrix technique. We used the well known transfer
matrix approach for self-avoiding walks introduced by Derrida \cite{D81}.
Due to the randomness, symmetry properties of the transfer matrix
disappear and in this way the size of the matrix is larger then in the
non-random case. Average over all bond
configurations could be performed on a Monte Carlo basis. At $1-p=p_c$.
the number of paths reaching the $x=t$-axis decreases as a power in $t$,
so that self-averaging cannot be used. Therefore a great number of
random configurations has to be used to obtain accurate statistics
(we used up to $4.10^6$ configurations). This limited our calculations
to the ranges $L \leq 9$ and $t \leq 100$.

For the small values of $L$ considered here no direct estimate
of $\omega$ or $D$ in the regime $t^{\zeta} \ll L$ was possible.
We were however able to get accurate estimates of $\omega/\zeta$ and
$D$ from the saturation value of $\overline{\Delta E}$ and the large
$t$ behaviour of $\overline{N(t)}$, respectively. Fig 2 shows a typical
result for $\overline{\Delta E(t)}$ for $L=9$ at $1-p=p_c$ with $q=1/2$.

Our results very clearly indicate $D=1$ in the whole region where $p_c <
1-p < 1$. For large values of $1-p$, $\omega/\zeta$ also clearly
goes to its DPRM value of 1/2 for large $L$.
Unfortunately in the region $p_c < 1-p < p_{cd}$ we are unable to draw
definite conclusions on the value of $\omega/\zeta$. The numerical
estimates are much larger then the DPRM-value, but decrease slowly with
increasing $L$. It is unclear whether the estimates will converge to
1/2 or to a higher value.
The results on the hierarchical lattice lead us to believe there is
no change in $\omega/\zeta$ at $p_{cd}$ and that the higher values
found here are due to crossover-effects. To solve completely the issue
on square lattice would require a computational effort beyond our
present possibilities.

Of particular interest are our determinations of $D$, and
$\omega/\zeta$ right at $(1-p)=p_c=1/2$.
Our results for D now clearly indicate that the optimal path becomes
a fractal with a dimension $D = 1.16 \pm 0.04$ with
much confidence. In view of the discussion and results in the previous section,
we should consider this as a determination
of $D_{\rm min}$ for the backbone of the $IIC$ in $2d$. This dimension
is not know exactly. Only a few numerical results exist,which are not
particularly consistent\cite{tanti}.
Our determination is not far from the independent ones found in the
literature for $D_{\rm min}$, and we think it should not be less accurate.

Finally, we extrapolated with reasonable accuracy $\omega/\zeta=1.02
\pm 0.06$. Because we are now in a fractal regime we get $\zeta=1$
and thus $\omega=1.02 \pm 0.06$, surprisingly consistent with the
values for hierarchical models considered in the previous
section.

As mentioned in the previous section, we did not succeed in getting
meaningful extrapolations for $\cal R$ with our FSS approach to self avoiding
paths.
Obviously the efficiency of the transfer matrix approach is
 much higher with directed paths. Much attention has been devoted
in the literature to  the case in which the directedness is oriented
along the diagonal of a square lattice \cite{HHb85,Kardar}. In such
condition it is
possible to construct a recursion relation which has a very simple structure.
In fact, due to the directedness, at each site only two bonds are merging
and we can reduce the optimization procedure
to just a choice between two possible energies.
For a given
realization of randomness,  the
energy of the optimal polymer configuration connecting the origin to all
 possible end points is easily computed by using iteratively the
recursion relation.

Although the number of possible paths grows  as $2^t$, the relatively
simple structure of the optimization procedure
makes this approach
extremely efficient and powerful. However, to the purpose
of this research paper, it is crucial to have
a distinction between the paths which minimize the energy and the ones
which have the minimal geometrical length. The square lattice with
directedness along the diagonal does not allow this distinction
 because all the
paths have the same length. We then must put the
directedness along one of the principal axis, e.g. $\hat{x}$, in order
to avoid this drawback. As for the undirected case the
paths have to be hosted in a strip of width $L$ along the direction
perpendicular to the directed one.

The recursion relation we must iterate becomes:

\begin{equation}
E_{\rm min}(y,x+1)={\rm \min}_{ 1 \leq k \leq L} \left\{
E_{\rm min}(k,x)+\sum_{i=k,y}E_{b \perp}(i,x)+E_{b \parallel}(y,x) \right\}
\label{eq:3_11}
\end{equation}
where $E_{b \perp}$ and $E_{b \parallel}$ represent the random energies which
are attributed to the bonds perpendicular or parallel to the oriented axis,
respectively.

Eq. (\ref{eq:3_11}) is more time and memory consuming than the one used
for the diagonal case,
  but its directed nature allows us to find the optimal solution
still in a polynomial time.

We have already pointed out for the undirected case that the
restriction of working with the size constraint measured by $L$ can
be partially overtaken from the possibility of estimating the scaling
exponents from the saturation values of $\Delta E$ by using equation
(\ref{eq:3_4}). However for this case we can still reach very high values of
the strip width $L$ (of the order of thousands), so we can also extract
directly
the critical indices $\zeta$ and $\omega$ through a standard log-log plot
of the suitable quantities. In all our calculations we tried to use both
methods in order to minimize statistical errors. Nevertheless these errors
are a little bit bigger than those usually found for the proper directed
problem in view of the shorter times  we are obliged to use.

Our first step was to investigate the best energy paths above
the percolation threshold of this model: $p_c = 0.555 \pm 0.002$\cite{Re81}.

As expected we found a quite good agreement with the DPRM exponents
$\omega=1/3$ and $\zeta=2/3$. By working in boxes up to $L=3000$
and $t=3000$ we found:

\begin{equation}
\zeta=0.65 \pm 0.04 \ \ \ \ \ \omega=0.34 \pm 0.02
\label{eq:3_12}
\end{equation}

Then we investigated the properties of the best
energy paths at the directed percolation threshold. This becomes quite
computationally intensive because a large fraction of the
configurations must be thrown away before a percolating cluster is found. We
could  obtain good statistics for samples up to $L=1000$ and $t=1000$
and we obtained:

\begin{equation}
\omega=0.51 \pm 0.03 \ \ \ \ \ \zeta=0.64 \pm 0.06
\label{eq:3_13}
\end{equation}

The results are very consistent with those obtained in ref. \cite{BK92}.
If we can not discriminate whether $\zeta$ is really different from
the DPRM value 2/3, we can certainly state that $\omega > 1/3$ as expected
from the calculation on the DHL.

In analogy with the work we did on the DDHL lattice we then tried to study
the difference between the best energetic path and the
shortest one. For each random configuration and for any time step
we computed the position of both  paths
and we stored
the difference $\Delta y$ between the two.
We have also defined
and analyzed a quantity ${\cal R}(t)$ which represents the probability
that after $t$ steps the optimal path
deviates from the shortest one.

 In this case the results do not
agree with those obtained for undirected paths on DDHL.
We have clearly found that the
quantity ${\cal R}(t)$ does not go to zero at $p_c$.
This conclusion is also confirmed by plots of
the ratio between the number of bonds in common between the two
kind of optimal paths and the total number of bonds as a function of $t$.
Moreover we have
seen that the average transversal displacement
( $\overline{\Delta y(t)}$) grows with $t$ following the law:

\begin{equation}
\overline{\Delta y(t)} \sim t^{\overline{\zeta}} = t^{0.67 \pm 0.07}
\label{eq:3_135}
\end{equation}.

This result and the fact that $\zeta$ can not be bigger than $\overline{\zeta}$
 support the conclusion that ${\overline{\zeta}}=\zeta$.

The above results show that a scaling exponent like $\rho$ can not be
defined at a threshold where the path remains self-affine.

\section{Conclusions}

In this work we carried on a systematic analysis of self-avoiding walks in
random environment with a fraction of forbidden bonds. We wanted
to establish up to which extent the removal of the directedness constraint
and the replacement of directed by isotropic percolation can lead
to new physics for the optimal  path solution.

Our analysis was partly based on hierarchical lattice RG calculations.
At the same time, in section 3 we set up a  new efficient FSS
scheme for the evaluation of optimal path properties
on 2d Euclidean lattice. Beyond the specific
results obtained here, this type of
scheme could have other
applications in the field. E.g., its generalization to finite temperature is
rather straightforward. One could also use it
for determining properties
of SAW in random environment referring to regimes
different from the one
treated in the present paper. For example, if the length
of the walk is properly controlled by a step fugacity, one expects that
the SAW can be made critical in the standard sense
of polymer statistics\cite{DM91}.
Working on DDHL, we also verified that simultaneous use of positive and
negative energy values in the distribution (\ref{eq:2_9}), upon varying $q$,
 allows
to identify a $T=0$ fixed point at the border
between the domains of attraction of our minimal length fixed point and
a similar one, dominated by maximal length paths on the backbone. This
border multicritical point turns out to have the scaling properties
expected for self-avoiding walks on percolation clusters on DDHL\cite{DM91}.

Thus our FSS methods could be used
in order to study problems like the SAW on
percolation clusters \cite{DM91,MH89,VK92}. An interesting methodological
indication we got here is that a transfer matrix
based FSS analysis of such problems
can give sensible results, especially as far as fractal dimensions are
concerned. Of course the full description of the crossover between
stretched minimal length and maximal length regimes would be extremely
time consuming in view of the necessity of testing many $q$ and
$L$ values.

Our calculation here show that the standard DPRM behaviour \cite{HHb85,KPZ} of
directed paths applies also
to the undirected case as long as $(1-p)>p_c$. Thus, the presence
of overhanging configurations does not alter the large scale behaviour of the
path, even for $p_c < (1-p)<p_{\rm cd}$, when backward
turns must be present in all path configurations. This is indication of
particular robustness of the standard DPRM behaviour.
A priori one could not exclude that
the directed percolation threshold could mark a transition to a different
self--affine scaling regime. Our results seem to rule out such possibility.

A further characterization of the
$(1-p) > p_c$ regime is given through the quantity ${\cal R}$, which measures
the probability  of detouchment of the optimal path from the minimal
length walk. In the self-affine regime ${\cal R}$ is always non-zero;
thus a non-zero fraction of optimal paths
deviates from the minimal length one.
For $(1-p)$ approaching $p_c$, ${\cal R}$ approaches zero, with a $\rho$
exponent close to unity on ``two dimensional"
hierarchical lattices.  The fact that ${\cal R}$ approaches zero
at threshold could be peculiar of cases in which the path geometry
crosses over from self-affine to fractal.
For directed paths at the directed threshold $p_{cd}$, ${\cal R}$
does not become zero, indicating that, even in threshold self-affine regimes,
the optimal path does not fully
stick on the minimal length path. As shown at the
end of the previous section, for directed paths at the directed
percolation threshold in 2d, while $\rho$ does not make sense,
it is possible
to describe how the distance between optimal and minimal length paths
diverges. We saw that this distance is still controlled by the usual
roughness exponent $\zeta$
of the paths.

At the undirected percolation threshold the
self-affine solution is replaced by a fractal one, with
fractal dimension equal to $D_{\rm min}$, the minimal length path dimension.
Our results, for both hierarchical and square lattice,
suggest an $\omega$ exponent  close to $1$ in this case.
$\omega=1$ strictly would be suggested by assuming validity of
the scaling relation (\ref{eq:2_4}) in the
limit of fractally rough path $( \zeta \to 1)$. On the other hand
we already know that eq. (\ref{eq:2_4}) does not hold at the directed
percolation threshold when directed paths are considered.
At the same time, in our case,
$\omega$ should be a direct manifestation of the statistical geometry of
minimal length paths on the $IIC$ backbone. The length fluctuations of the
shortest path within the backbone are enough to
determine the observed energy fluctuations. Thus,
in the fractal regime of the
optimal path, there is no appreciable role
played by the bond energy
probability distribution in the
sense of determining a non-trivial
stable law for minimal energy fluctuations. Minimal
length fluctuations, merely due
to dilution, are the dominating source of
energy variance.  Geometry alone seems to fully control the optimal path
scaling at the percolation threshold.

We could also verify that, like in the self--affine regimes, at
$1-p=p_{c}$, $\omega$ also
characterizes the subleading correction to
the dominant behaviour of
$\overline{E_{\rm min}}$ and qualifies as the only scaling dimension for
energy according to eq. (\ref{eq:2_11}).

Our determination of $D_{\rm min}$ on square lattice
seems rather consistent if compared
with others in the literature\cite{Kinzel83}.
New interest in this kind of percolation cluster dimension has been stimulated
 most
recently by the realization that it controls the dynamics of directed
percolation
interface depinning\cite{tanti}.

The exponent $\omega$ we determined, besides constituting,
like $\rho$, an important characterization
of the fractal regime of the optimal path,
reflects an interesting, and, so far, unexplored,
property of the backbone geometry. It would
be interesting to have determinations of it in
higher dimensions. Unfortunately, calculations
of $\omega$ on $3d$ Euclidean lattice are
definitely out of reach with our methods.

In order to get a preliminary indication on higher $d$ situations,
we decided to try
an RG calculation for a hierarchical lattice, which,
after removing the diagonals, coincides with what can be taken
as a surrogate of $d>2$ lattice for directed problems (See Fig 3 )\cite{DG89}.

The thresholds for undirected and directed bond percolation
on this lattice are $p_c=0.2760\ldots$ and $p_{\rm cd}=0.3893\ldots$,
 respectively.
 By the methods of section 2 we obtained for optimal self
avoiding paths on this lattice: $D=1.185 \pm 0.005 $ very close to the exact
$D_{min}$, and $\omega=1.16 \pm 0.05$, definitely larger than $1$.

A direction in which it is conceivable to
extend the present investigation is that of
including nonzero temperature effects. As already remarked
above, this can be done by our
methods. E.g., in the hierarchical cases, in the iteration
strategy the minimal energy distribution would
be replaced by a distribution for
the full partition function of self-avoiding walks,
 $Z=\sum_{w}\exp{(-E_W/k_bT)}$.

For the models described in the previous sections, we expect asymptotic
distributions
controlling the physics at large scales to
be the same as those already experienced at $T=0$. Strong
disorder, $T=0$ distributions always
determine the scaling, at least in the stretched phase. On the
other hand, for $d \geq 3$ on Euclidean lattices, the $T=0$ strong disorder
regime
is known to extend  only up to some critical temperature,
marking a transition to a high $T$ regime where disorder
becomes irrelevant \cite{DGX} ($\omega=0, \zeta=1/2)$.
Possible candidates to show this transition, and, at the same time, the
crossover to
fractal regime under dilution, are hierarchical
models of the type described in Fig. 3.
It would be interesting, though quite laborious, to
establish the effect of directedness removal and dilution
on the finite temperature transition, in particular at the fractal thresholds.

\newpage

\centerline{\bf Figure Captions}

\vskip 3truecm

{\bf Fig 1}. Iterative construction of the four hierarchical lattices:
1a) The DDHL lattice
(the bonds are numbered and the paths to be considered in the iterative RG
calculation are:$1,3;\  1,5,4;\  2,4;\  2,5,3.$); 1b)  The DHL lattice;
1c) A hierarchical lattice with $p_c=0.70830\ldots$ and $p_{cd}=0.72578\ldots$;
1d) A hierarchical lattice with
$p_c=0.54984\ldots$ and $p_{cd}=0.55194\ldots$.

\vskip 3truecm
{\bf Fig. 2}. Result of transfer matrix calculation
for $\overline{\Delta E(t)}$
as a function of t for $L=9,1-p=q=1/2$.

\vskip 3 truecm

{\bf Fig. 3}.  Iterative construction (first step)  of a ``higher dimensional"
hierarchical lattice.

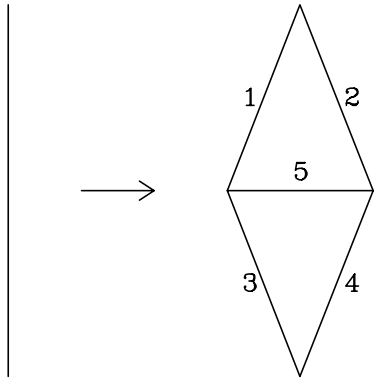
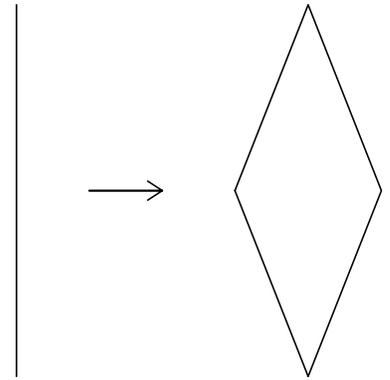
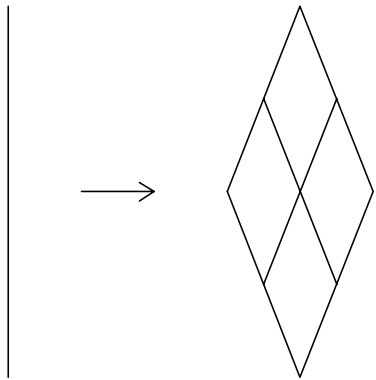
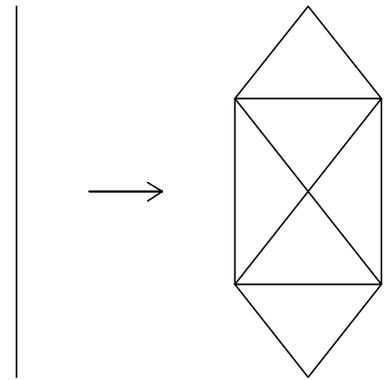
\end{document}